\documentstyle[preprint,aps,psfig]{revtex}
\begin{document}
\draft

%\twocolumn[\hsize\textwidth\columnwidth\hsize\csname @twocolumnfalse\endcsname

\title{Molecular crystal approach for $\pi$-conjugated polymers: from PPP
Hamiltonian to Holstein model for polaron states}

\author{St\'{e}phane Pleutin$^{\dag *}$ and Jean-Louis Fave$^{*}$}
\address{
$^{\dag}$Max-Planck-Institut f\"ur Physik Komplexer Systeme, N\"othnitzer
Stra\ss e 38, D-01187 Dresden\\
$^{*}$Groupe de Physique des Solides, 2 place Jussieu, 75251 Paris cedex 05}

\date{\today}
\maketitle

\begin{abstract}
Starting from the $\pi$-electron Pariser-Parr-Pople (PPP) Hamiltonian
which includes both strong electron-phonon and electron-electron
interactions, we propose some strongly correlated wave functions of
increasing quality for the ground
state of conjugated polymers. These wavefunctions are built by combining
different finite sets of local configurations extended at most over two
nearest-neighbour monomers. With this picture, the doped case with one
additional particle is expressed in terms of
quasi-particle. Thus, the polaron formation problem goes back to the study
of a Holstein like model.
\end{abstract}

%\pacs{ PACS Numbers: 71.10.-w, 71.20.Rv, 71.38.+i}
\vskip2pc
\narrowtext
\newpage

\section{Introduction}

The nature of the first excited states of conjugated polymers is an important
and still unsolved question in condensed matter sciences \cite{excitations}. Knowing if they are
band to band excitations or exciton states, if polarons, bipolarons or
solitons are
stable quasiparticles in the doping case, are fundamental issues for the
understanding of the electronic properties of these compounds.

The low-lying excited states are supposed to be suitably described by
the well-known $\pi$-electron Pariser-Parr-Pople (PPP) Hamiltonian. This
model
Hamiltonian takes into
account both strong electron-phonon and electron-electron interaction terms
yielding only exact numerical solutions for the smallest
oligomers\cite{chandross}. For the
thermodynamic limit, the situation remains unclear since the calculations for the
ground state and the excited states, including
electron correlations, are uneasy to achieve and some drastic
approximations are needed \cite{revue}.

However, a first qualitative understanding of this complicated physics
can be done by the use of some simplified Hamiltonian. For instance the
Rice-Gartstein's molecular exciton model\cite{rice}, similar to previous works \cite{excitonic}, is useful for a qualitative
description of the linear absorption of conjugated
polymers. On the other hand, the molecular
Holstein's model gives a simplified picture of the polaron
states\cite{holstein,bussac}.

Recently, an approximate scheme to build the ground and the first
excited states has been proposed\cite{pleutin}. With this method,
starting from the PPP Hamiltonian, one reaches a Rice-Gartstein like
model; the excitations relevant for linear absorption are
then easy to obtain and the results are comparable with those from more tedious
methods \cite{yu}. In this paper, we will show that the same procedure permits to derive
formally, from the very same PPP Hamiltonian, the simple molecular
Holstein's model for the polaron state in conjugated polymers. Polarons
are thought to be important for the understanding of the charge
transport in these compounds, and the possibility to study these
non-linear states at a correlated level in an easy formalism is needed
and valuable. 

We choose a simple dimerized linear chain as an effective model for conjugated
polymers; this chain is characterized by $r_{d}$ and $r_{s}$, the
double and single bond lengths, respectively. Extending our method
to a realistic geometry is straightforward but
the essential of the physics is reached with this simplified
picture. Let us now briefly introduce the Pariser-Parr-Pople
Hamiltonian which is our starting point

\begin{equation}
\label{PPP}
H_{PPP}=-\sum_{n,\sigma}t_{n,n+1}(c^{\dag}_{n,\sigma}c_{n+1,\sigma}+
c^{\dag}_{n+1,\sigma}c_{n,\sigma})+
\frac{1}{2}\sum_{n,m,\sigma,\sigma'}V_{n,m}
(c^{\dag}_{n,\sigma}c_{n,\sigma}-\frac{1}{2})
(c^{\dag}_{m,\sigma'}c_{m,\sigma'}-\frac{1}{2})
\end{equation}
where $c^{\dag}_{n,\sigma}$, ($c_{n,\sigma}$) is the creation
(destruction) operator of an electron in site {\sl n} with spin $\sigma$;
$t_{n,n+1}$ is the hopping term which includes via the electron-phonon
interaction a linear dependence upon the length of the bond
({\sl n},{\sl n+1})\cite{revue,tavan}. In the case
of a dimerized linear chain, this dependence gives two distinct hopping
integrals $t_{d}$ and $t_{s}$ for the double and the single bonds
respectively ($\mid t_{d} \mid > \mid t_{s} \mid$); they could be
written as $t_{d/s}=t_{0}(1\pm \frac{\alpha}{2t_{0}}\delta)$ where
$t_{0}$ is the hopping integral without dimerization, $\alpha$ is the
electron-phonon interaction and $\delta$ is a measure of the
dimerization giving the difference of the lengths of single and double
bonds\cite{tavan}.
The Coulomb term is parametrized following Ohno, where the effect of the
$\sigma$ electrons is considered via a phenomenologic screening,
$V_{n,m}=\frac{U}{\sqrt{1+0,6117r^{2}_{n,m}}}$ where $r_{n,m}$ is the distance
(in $\AA$) between two electrons localized on site {\sl n} and {\sl m} \cite{ohno}. We also
write this term as $V(r_{n,m})\equiv V_{n,m}$ and $V=V(r_{d})$ for convenience.

In view to link up the PPP Hamiltonian and the molecular
crystal models, the Rice-Gardstein and Holstein models, we choose the monomer self-consistent orbitals as basis
functions - this is the so-called exciton-basis \cite{chandross}. This choice is of course led by the dimerization. In our case,
the monomers are the double-bonds and their
self-consistent orbitals are associated with the following creation
(destruction) operators for the bonding and anti-bonding orbitals:
$B^{(\dag)}_{n,\sigma}=\frac{1}{\sqrt{2}}(c^{(\dag)}_{2n,\sigma}+
c^{(\dag)}_{2n+1,\sigma})$ and $A^{(\dag)}_{n,\sigma}=\frac{1}{\sqrt{2}}
(c^{(\dag)}_{2n,\sigma}-c^{(\dag)}_{2n+1,\sigma})$; here $n$
indexes the double bonds.

With this specific choice of local basis operators, the electronic
configurations are built by combining different kinds of local
configurations (LC)\cite{chandross,these}. In order to get a tractable model,
we truncate the Hilbert space by choosing a small set of different LC which
will be the elemental building blocks for the electronic
configurations \cite{these}. These LC are the so-called generative local
configurations (GLC) in \cite{pleutin}. 

We may notice that this
method shows some similarities with the Valence Bond method used efficiently for
the studies of oligomers \cite{soos} but with the important
difference that atomic sites are replaced by monomer units with
internal electronic structure (double bonds here). The configurations
build from GLC are diagonal with respect to the hopping term $t_{d}$
on contrary to the Valence Bond configurations which are diagonal
with respect to the Coulomb term. Each GLC is a set of
several Valence Bond diagrams, chosen to be the adequate ones for a reasonable
description of polymer states.

In this work, we first improve the proposed ground state of
ref. \cite{pleutin}
by enlarging the set of electronic configurations used to describe
it (section II). Second, we consider the case with an extra
electron on the chain and show that, if one authorizes small lattice
distortions around the extra particle, our treatment allows to reach
quite naturally a Holstein like model but expressed in terms of
many-body particle states (section III).

\section{The ground state}

\begin{figure}
\centerline{\psfig{figure=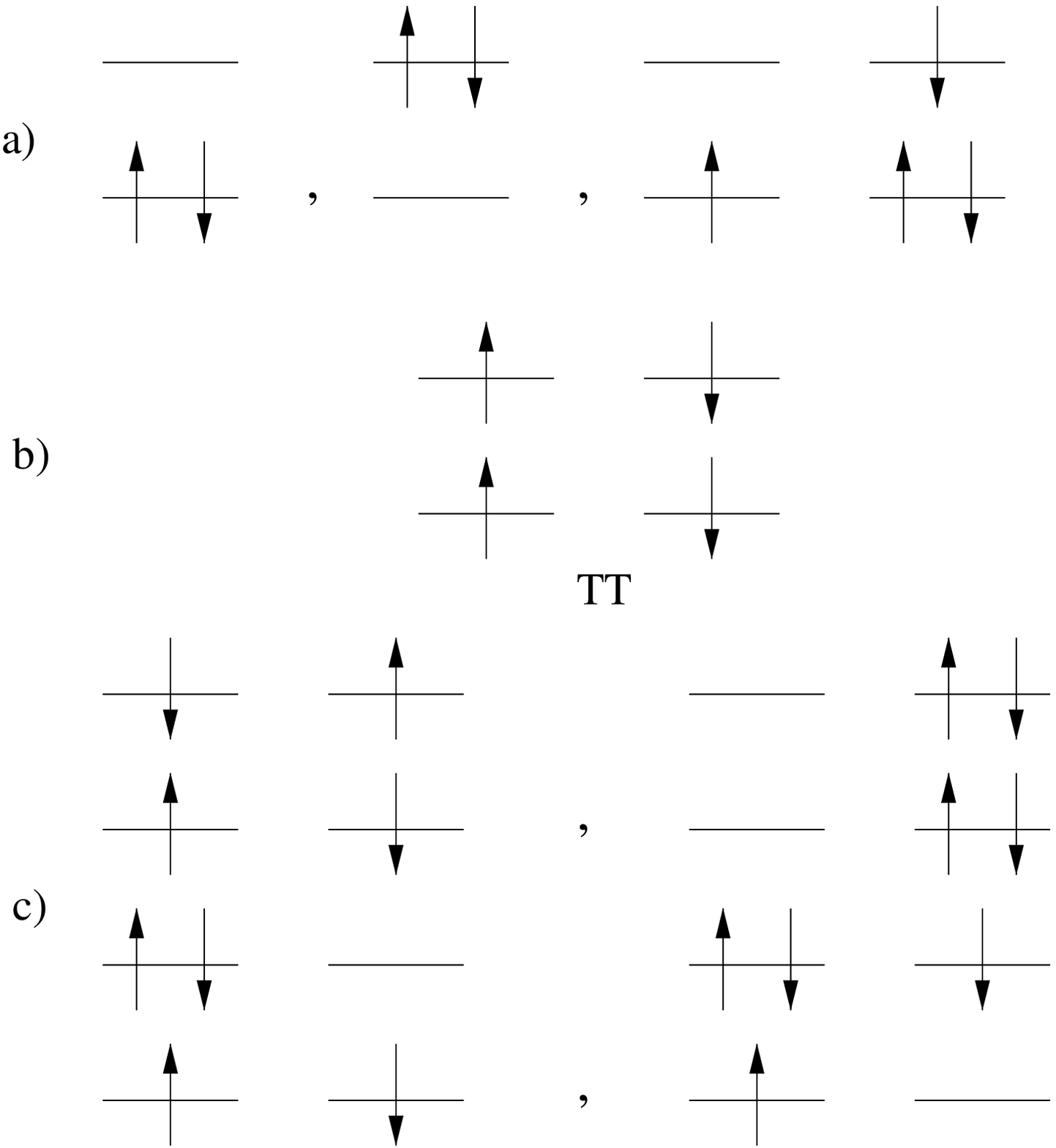,width=8cm}}
\caption{Generative-Local-Configurations (GLC) selected to build the ground
state wave functions. The set of GLC (a) defines the model I, the set of GLC (a+b) defines the model II and the whole
set of GLC (a+b+c) defines the model III.}
\label{LCGS}
\end{figure}

We keep as GLC for the ground state the LC which appear the most
relevant in calculations performed on small oligomers\cite{chandross}. In ref
\cite{pleutin}, only three LC were considered; they are named F-LC, D-LC and
Ct$^{-}_{1}$-LC and are schematically represented in figure
(\ref{LCGS}.a). This approximation could appear rather abrupt, but it
is sufficient to get a correct qualitative picture of the linear
absorption spectra as it was shown in \cite{pleutin}; moreover, even at
this level of approximation, the results are quantitatively comparable
with the results of more tedious calculations \cite{yu}. In this work, we propose
some natural improvements to this first approximation by extending the set of
GLC. 

In a first improvement, we add to the previous set of GLC, the so-called
Triplet-Triplet LC, TT-LC, shown in figure
(\ref{LCGS}.b), where two nearest-neighbour (n.n.) localized triplets are
combined into a singlet. Together with the three first LC, they are
the major constituents of the ground state wave
function in small cluster calculations\cite{chandross}. In a second
improvement, we enlarge again the set of GLC by including in it the LC which
interact directly with the four previous selected ones (figure\ref{LCGS}.c).

In the following, only the first case is treated explicitely. We
develop in full detail our proposed way to get the ground state wave
function with the four selected GLC. The case with the complete set of LC
represented in figure (\ref{LCGS}) can be treated following the same
scheme; only the obtained results are then given.

First, we introduce the four GLC, their associated creation operators
and their energies.
\begin{itemize}

\item The named F-LC is associated with the creation operator
\begin{equation}
F^{\dag}_{n}=B^{\dag}_{n,\uparrow}B^{\dag}_{n,\downarrow}
\end{equation}This define the lowest LC in the range of parameter of
interest; therefore we choose as reference state 
\begin{equation}
\mid 0>= \prod_{n}F^{\dag}_{n}\mid Vacuum>
\end{equation}
where $\mid Vacuum>$ denotes the state without any $\pi$
electron. The state $\mid 0>$ is the ground state considered in the molecular
crystal approaches \cite{rice,excitonic}; there, the linear dimerized chain is
simply identified to an one-dimensional crystal of ethylene without any
electronic correlations.

With respect to $\mid 0>$, $F^{\dag}_{n}=I^{\dag}_{n}$ which
is simply the identity operator. In the following, all the creation operators
and the energies are defined with respect to $\mid 0>$.

\item The named D-LC is associated with the creation operator
\begin{equation}
D^{\dag}_{n}=A^{\dag}_{n,\uparrow}A^{\dag}_{n,\downarrow}
B_{n,\uparrow}B_{n,\downarrow}
\end{equation}and with energy given by $\epsilon_{d}=4t_{d}$. 

The F and D-LC describe the dynamics of
the $\pi$-electrons coupled by pairs into each monomer: the two electrons are
independent in F-LC, whereas D-LC introduces intramonomer electronic
correlation. In the strong dimerization limit, these two LC are
sufficient to give a good approximation of the ground state; the system is
then very close to a true molecular crystal. For small or
intermediate dimerization, it is however necessary to consider more extended
LC or, in other words, some fluctuations around the molecular crystal
limit. This is done by introducing two more LC extended over two
n.n. monomers.

\item The named Ct$^{-}_{1}$-LC is associated with the creation operator
\begin{equation}
\label{ctLC}
Ct^{\dag}_{n}=\frac{1}{2}(A^{\dag}_{n+1,\uparrow}
B_{n,\uparrow}+A^{\dag}_{n+1,\downarrow}B_{n,\downarrow}-
A^{\dag}_{n,\uparrow}B_{n+1,\uparrow}-A^{\dag}_{n,\downarrow}B_{n+1,\downarrow})
\end{equation}and with energies given in the case of a linear dimerized chain by
$\epsilon_{ct}=2t_{d}+V-\frac{1}{4}(V(r_{s})+2V(r_{s}+r_{d})+V(2r_{d}+r_{s}))$.
The last term, in bracket, is the attractive interaction between the
electron and the hole
due to the long-range part of the Ohno potential.

The Ct$^{-}_{1}$-LC introduces n.n.
intermonomer charge fluctuations, reproducing the conjugation
phenomenon in a minimal way.

\item Last, the named TT-LC is associated with the creation operator
\begin{equation}
\begin{array}{c}
TT^{\dag}_{n}=\frac{1}{\sqrt{3}}(A^{\dag}_{n,\uparrow}B_{n,\downarrow}A^{\dag}_{n+1,\downarrow}B_{n+1,\uparrow}+A^{\dag}_{n,\downarrow}B_{n,\uparrow}A^{\dag}_{n+1,\uparrow}B_{n+1,\downarrow}+
\frac{1}{2}(A^{\dag}_{n,\uparrow}B_{n,\uparrow}A^{\dag}_{n+1,\uparrow}B_{n+1,\uparrow}+\\A^{\dag}_{n,\uparrow}B_{n,\uparrow}A^{\dag}_{n+1,\downarrow}B_{n+1,\downarrow}+A^{\dag}_{n,\downarrow}B_{n,\downarrow}A^{\dag}_{n+1,\uparrow}B_{n+1,\uparrow}+A^{\dag}_{n,\downarrow}B_{n,\downarrow}A^{\dag}_{n+1,\downarrow}B_{n+1,\downarrow}))
\end{array}
\end{equation}and with energy given by $\epsilon_{tt}=4t_{d}-(U-V)$. 

In this LC, two Triplets appearing in n.n. monomers are
combined into a singlet (figure(\ref{LCGS}.b)). It was shown to be
important for the first time in the work of Schulten and Karplus
\cite{schulten} where it was recognized as a major
constituent of the low-lying excitations, the famous $2A_{g}^{-}$
state, optically forbidden. In the ground state, which is our interest
here, the importance of this LC can be
comparable to the D-LC one \cite{chandross}. 
\end{itemize}

We may notice that a similar treatment for the PPP Hamiltonian was proposed a
few years ago to study the spin-charge separation mechanism in the limit of
strong dimerization \cite{mukho}.

With our choice of four GLC, all possible electronic configurations
are then build up. They are characterized by the number of D, Ct$^{-}_{1}$ and
TT-LC, $n_{d}$, $n_{ct}$ and $n_{tt}$ respectively, and by the
positions of these different GLC. The positions of the D, Ct$^{-}_{1}$ and TT-LC
are labelled by the coordinates $z(k)$ ($k=1,..,n_{d}$), $y(j)$
($j=1,..,n_{ct}$) and $x(i)$ ($i=1,..,n_{tt}$) respectively. The necessary
non-overlapping condition between LC is supposed to be fulfilled all along the
paper - the LC behave as hard core bosons. The electronic configurations are then expressed as

\begin{equation}
\label{espacemodel}
\mid x(1),...,x(n_{tt}),y(1),...,y(n_{ct}),z(1),...,z(n_{d})>=\prod_{i=1}^{n_{tt}}\prod_{j=1}^{n_{ct}}\prod_{k=1}^{n_{d}}TT^{\dag}_{x(i)}Ct^{\dag}_{y(j)}D^{\dag}_{z(k)}\mid
0>
\end{equation}The GLC are all neutral local configurations, therefore the
energy of (\ref{espacemodel}) is independent of the
relative positions between LC and entirely determined by the
number of each GLC.
\begin{equation}
\label{energiemodel}
E(n_{tt},n_{ct},n_{d})=n_{tt}\epsilon_{tt}+n_{ct}\epsilon_{d}+n_{d}\epsilon_{d}
\end{equation}

At this point, we have to mention an incorrect statement in \cite{pleutin}
where it is saying that the energy of the configurations made of F, D and
Ct$^{-}_{1}$-LC depends on the relative positions of the Ct$^{-}_{1}$-LC. This 
statement is actually wrong, however, this simplification goes in
favor of our treatment (indeed, it was not possible to do calculations with
this statement and finally the energy (\ref{energiemodel}) was also adopted in \cite{pleutin}).

The way we choose to diagonalize the PPP Hamiltonian in the reduced Hilbert space spanned by the electronic configurations
(\ref{espacemodel}) follows from \cite{pleutin}. First, we reorganize
the configurations (\ref{espacemodel}). We make linear combinations
from the states with $n_{d}$ D-LC, $n_{tt}$ TT-LC localized at sites
$x(1),...,x(n_{tt})$ and $n_{ct}$ Ct$^{-}_{1}$-LC localized at sites
$y(1),...,y(n_{ct})$. Since we are interested, at the end of the day,
only by the lowest state in energy (the ground state), we can consider
only the linear combinations of highest symmetry
\begin{equation}
\label{Excocorr}
\mid
x(1),...,x(n_{tt}),y(1),...,y(n_{ct}),n_{d}>=\frac{1}{\sqrt{C_{n_{d}}^{N-2(n_{tt}+n_{ct})}}}\sum_{\{z(k)\}}\prod_{k=1}^{n_{d}}D^{\dag}_{z(k)}\prod_{i=1}^{n_{tt}}\prod_{j=1}^{n_{ct}}TT^{\dag}_{x(i)}Ct^{\dag}_{y(j)}\mid
0>
\end{equation}where the summation is carried over the
$C_{n_{d}}^{N-2(n_{tt}+n_{ct})}$ possible configurations. The energy
of these combinations is still given by
(\ref{energiemodel}).

The states (\ref{Excocorr}) interact between them by the following term
\begin{equation}
\label{interacD}
\begin{array}{c}
<x(1),...,x(n_{tt}),y(1),...,y(n_{ct}),n_{d}\mid
H_{PPP}\mid
x(1),...,x(n_{tt}),y(1),...,y(n_{ct}),n_{d}+1>=\\
\sqrt{(n_{d}+1)(N-2(n_{tt}+n_{ct})-n_{d})}\frac{U-V}{2}
\end{array}
\end{equation}

The tri-diagonal matrix, where the diagonal part is given by
(\ref{energiemodel}) and the off-diagonal part by (\ref{interacD})
can be divided into sub-matrices characterized by $n_{ct}$ localized
Ct$_{1}^{-}$-LC and $n_{tt}$ localized TT-LC but with a variable
number of D-LC, $n_{d}$ ($n_{d}=0,..., 2(n_{ct}+n_{tt})$); these sub-matrices can be
separately diagonalized and it is easy to show that
the resulting lowest states are given by the following expression

\begin{equation}
\label{diago1}
\begin{array}{c}
\mid
x(1),...,x(n_{tt}),y(1),...,y(n_{ct})>^{c}=\sum_{n_{d}=0}^{N-2(n_{tt}+n_{ct})}a^{N-2(n_{tt}+n_{ct})-n_{d}}b^{n_{d}}\sqrt{C_{n_{d}}^{N-2(n_{tt}+n_{ct})}}\\
\mid
x(1),...,x(n_{tt}),y(1),...,y(n_{ct}),n_{d}>
\end{array}
\end{equation}with energy expressed as

\begin{equation}
\label{energiecor}
E^{c}(n_{tt},n_{ct})=n_{tt}\epsilon_{tt}+n_{ct}\epsilon_{ct}+(N-2(n_{tt}+n_{ct}))\epsilon_{c}
\end{equation}where
\begin{equation}
\label{Ec}
\epsilon_{c}=2t_{d}-\frac{1}{2}\sqrt{16t_{d}^{2}+(U-V)^{2}}
\end{equation}The coefficients $a$ and $b$ of (\ref{diago1}) are given by
$a=\frac{U-V}{\sqrt{4\epsilon_{c}^{2}+(U-V)^{2}}}$ and
$a^{2}+b^{2}=1$. With these expressions, the double bonds free of
Ct$_{1}^{-}$- and TT-LC are correlated independently. The
upper-script, $c$,
in (\ref{diago1}) is for correlated. $\epsilon_{c}$ is called
intramonomer correlation energy.

The next step toward the evaluation of the ground state is to retain,
among all the states resulting from the previous sub-diagonalizations, only
the lowest ones given by (\ref{diago1}). This approximation is well
justified since the energy difference between these states and the
corresponding lowest excited ones is given by the quantity
$\sqrt{16t_{d}^{2}+(U-V)^{2}}$ which is rather high for usual
parameters with a value around $10eV$. We then reorganized the
states (\ref{diago1}) into collective
excitations of highest symmetry
\begin{equation}
\mid n_{tt},n_{ct}>^{c}=[C_{n_{tt}+n_{ct}}^{N-n_{tt}-n_{ct}}C_{n_{tt}}^{n_{tt}+n_{ct}}]^{-\frac{1}{2}}\sum_{\{x(i),y(j)\}}\mid
x(1),...,x(n_{tt}),y(1),...,y(n_{ct})>^{c}
\end{equation}still associated with the energy (\ref{energiecor}) and
where the summation runs over the
$C_{n_{tt}+n_{ct}}^{N-n_{tt}-n_{ct}}C_{n_{tt}}^{n_{tt}+n_{ct}}$
possible configurations. The
ground state is then expressed as a linear combination

\begin{equation}
\label{wfgs}
\mid GS>=\sum_{n_{tt},n_{ct}}X_{n_{tt},n_{ct}}\mid n_{tt},n_{ct}>^{c}
\end{equation}where the coefficients $X_{n_{tt},n_{ct}}$ are determined by
solving the following secular equation

\begin{equation}
\label{equationgs}
\begin{array}{c}
I(n_{tt},n_{ct}-1)X_{n_{tt},n_{ct}-1}+
(n_{tt}\epsilon_{tt}+n_{ct}\epsilon_{ct}-2(n_{tt}+n_{ct})\epsilon_{c}-E)X_{n_{tt},n_{ct}}+
I(n_{tt},n_{ct}+1)X_{n_{tt},n_{ct}+1}+\\\
[n_{tt}(n_{ct}+1)]^{\frac{1}{2}}n_{tt}\frac{\sqrt{3}}{2}t_{s}X_{n_{tt}-1,n_{ct}+1}+
[n_{ct}(n_{tt}+1)]^{-\frac{1}{2}}n_{ct}\frac{\sqrt{3}}{2}t_{s}X_{n_{tt}+1,n_{ct}-1}=0
\end{array}
\end{equation}where

\begin{equation}
\label{I}
I(n_{tt},n_{ct})=\sqrt{(n_{ct}+1)\frac{(N-2(n_{tt}+n_{ct})-1)(N-2(n_{tt}+n_{ct}))}{N-n_{tt}-n_{ct}}}a^{2}t_{s}
\end{equation}

The equation (\ref{equationgs}) is not solvable with the interaction
term (\ref{I}). Next, and last, we do an approximation on the term
$I(n_{tt},n_{ct})$ by assuming

\begin{equation}
\label{approx}
I(n_{tt},n_{ct}) \simeq
\sqrt{(n_{ct}+1)(\frac{N-1}{3}-n_{tt}-n_{ct})}\sqrt{3}a^{2}t_{s}
\end{equation}This is a very good approximation of (\ref{I}), if the number of
GLC extended over two monomers, $n_{2}=n_{tt}+n_{ct}$, is
small\cite{pleutin}. Consequently, this treatment will be justified if in the final wave function, the
most important configurations are the ones with a small value of
$n_{2}$; this is actually the case as it can be seen from the work of
ref. \cite{pleutin} and as it appears, at the end of the day, in this study.

With the last simplification, the problem is mapped onto $(N-1)/3$
independent three-level systems. One write

\begin{equation}
\begin{array}{c}
X_{n_{tt},n_{ct}}=\sqrt{C_{n_{tt}+n_{ct}}^{\bf{E} ((N-1)/3)}C_{n_{tt}}^{n_{tt}+n_{ct}}}y_{n_{tt},n_{ct}}\\
\mbox{with} \left \{
\begin{array}{c}
y_{n_{tt},n_{ct}}/y_{n_{tt}+1,n_{ct}}=\gamma \\
y_{n_{tt},n_{ct}}/y_{n_{tt},n_{ct}+1}=\zeta
\end{array}\right.
\end{array}
\end{equation}where $\bf{E}$ takes the integer part, $\gamma$ and
$\zeta$ are real constants to be determined. Inserting this definition in
(\ref{equationgs}) and after some algebraic manipulations one finds that
the problem goes back to calculate the lowest eigenvalue, $\epsilon$, of the
following 3 by 3 matrix

\begin{equation}
\label{3levels}
\left (
\begin{array}{c}
0 \quad \sqrt{3}a^{2}t_{s} \quad \quad \\
\sqrt{3}a^{2}t_{s} \quad \epsilon_{ct}-2\epsilon_{c} \quad
\frac{\sqrt{3}}{2}t_{s} \\
\quad \quad \frac{\sqrt{3}}{2}t_{s} \quad \epsilon_{tt}-2\epsilon_{c}
\end{array} \right )
\end{equation}The ground state energy is then simply divided into two
different components

\begin{equation}
\label{EGS}
E_{GS}=N \epsilon_{c}+\frac{N-1}{3}\epsilon
\end{equation}The first part is the intramonomer
correlation energy defined by the first subdiagonalization; it is obtained
by correlating independently the $N$ double bonds. The second part
is the intermonomer fluctuation energy defined by
the second subdiagonalization; it is obtained by considering $(N-1)/3$
identical and independent effective three level systems defined by the
matrix (\ref{3levels}). Finally, the ground state wave
function is clarified by the following two equations

\begin{equation}
\gamma=\frac{a^{2}t_{s}}{\epsilon} \quad , \quad \zeta=\frac{2a^{2}}{\sqrt{3}}\frac{\epsilon-\epsilon_{tt}}{\epsilon}
\end{equation}
The resulting wave function contains, as the energy, two
different kinds
of components: the first ones localize electrons by pairs in the
double bonds; the second ones introduce n.n.
intermonomer fluctuations, charge fluctuations by means of Ct$_{1}^{-}$-LC and 
spin fluctuation by means of TT-LC.

The ground state proposed above may be easily improved by adding new
local configurations extended over two n.n. double
bonds. For example, one can include the whole LC represented in the
figure (\ref{LCGS}); the LC of (\ref{LCGS}.c) are the ones directly coupled to
the others. The strategy is then the same. First, one takes
care of the intramonomer correlation; second, one builds the collective
excitations of highest symmetry; third, one approximates the part of
the resulting interaction connecting configurations which differ by only one LC
extended over two monomers in the way of (\ref{approx}). The problem is
then equivalent to consider $(N-1)/3$ independent seven-level systems; $\epsilon$ is then the
lowest eigenvalue of the associated 7 by 7 matrix.

In order to test our assumptions from which we propose several ground state
wave functions in the form of (\ref{wfgs}), we do comparisons, first, for the Su-Schrieffer-Heeger (SSH) model.
For this model, similar to (\ref{PPP}) but without the complicated Coulomb
term\cite{ssh}, the exact result is well known
\cite{ssh,salem}. We compare this result with successively the results given by the
model ground state of ref \cite{pleutin}, the hereafter so-called model I, the
one with in addition the
TT-LC, the model II, and, last, the model with all the GLC represented
in figure (\ref{LCGS}), the model III. We make comparison in function of the
dimerization parameter $x=\frac{\alpha}{2t_{0}}\delta$. The results
are shown in Table I where the percentage of the exact energy for
our successive approximations are given. For $x=1$, the
case of complete dimerization, the
three models give obviously the exact result. For $x=0$, the case
without dimerization, one gets around 92$\%$ of the total
energy. A-priori in this limit, one would expect less accurate results since
the charge fluctuations of longer range than one play
a role; they contribute in fact only in the missing 6$\%$. For
$x=0.15$, a value often attributed to the polyacetylene, one gets
around 97$\%$ of the total energy. In conclusion, our approximation seems rather
good for realistic cases, within this independent electron model.

Next, we do also comparisons for the Hubbard model which is well known
to be
exactly solvable in one dimension \cite{lieb}; this is the model
(\ref{PPP}) with $\alpha=0$ and where only the on-site
electron-electron interaction, $U$, is
retained. For $U=0$, one gets the SSH model without dimerization for
which we obtained around 92$\%$ of the total energy (see Table
I). Starting from
these values, the agreement monotonically decreases when
$U$ increases to finally get for infinite $U$, between 77$\%$ and
79$\%$ of the total energy, depending on the model (I, II or III) under
consideration. This discrepancy shows that important LC are missing
especially in the strong $U$ limit; for instance, it is easy to see,
just by energetic considerations, that, for large enough $U$, the TTT-LC, which is singlet made by three
localized triplets, the TTTT-LC, which is singlet made by four localized
triplets, and so on, may become important for the ground state wave
function. With our specific choice of basis set
completely localized on the double bonds, the dimerization parameter,
$x$, is crucial; the more it will be important, the more our treatment
will be relevant to become exact for a complete dimerization. In the
Hubbard model, the dimerization is simply missing. If $\alpha \ne 0$,
the energy of the LC made from localized triplets increases making our
approximations more and more reasonable.

Last, we do comparison for the so-called extended Peierls-Hubbard
model; this is the model (\ref{PPP}) but with only the Hubbard term,
$U$, and the n.n. interaction $V$, with the assumption that
$V=V(r_{d})=V(r_{s})$\cite{eric1}. On the contrary to the two previous
models, this model is not integrable, also, we do comparisons with
calculations performed with the Density-Matrix Renormalization Group
(DMRG) technique \cite{white}; a very
recent review of the advances related to this method may be found in
\cite{dmrg}.  The
DMRG calculations have been done by E. Jeckelmann \cite{eric2}
following the method developed in ref \cite{eric1}. We compare our
approximate results with an extrapolation of the energy per unit
cell made from calculations for different lattice lengths up to two hundreds
double bonds. The calculations are performed for a reasonable choice of
parameters, $U=4t_{0}$ and $V=t_{0}$. The results for several values
of the dimerization parameter are confined in table II. We see that the
errors are always less than 20$\%$ and are around 13 - 10$\%$ for
realistic parameters. In our opinion, the agreements obtained here are
satisfactory considering the relative simplicity of the wave functions
proposed in this work. Moreover, with these approximate wave functions, some
analytical insights are now possible which is very new in this range of
parameters, appropriated for conjugated polymers.

We do not compare for the moment our results with calculations made
for the complete PPP Hamiltonian. Nevertheless,
since the
remaining long range terms of the Coulomb potential are of smaller
importance than the other terms of the Hamiltonian, one can reasonably
expect only small quantitative changes in the results obtained with the extended
Peierls-Hubbard model by using the full PPP Hamiltonian.

Before closing this section, we may say that our wave functions are
not variational since, in the
way we choose to diagonalize the model, we do two successive
sub-diagonalizations with some approximations. However, it is possible to build 
some variational wave-functions very similar to (\ref{wfgs}). By the way, works are already done to propose a variational version of the
model II \cite{rva1} and other are in progress for the model III
\cite{rva2}. In other way, a very efficient Matrix-Product-Ansatz is also
proposed in \cite{mp}. Compared to the work developed in \cite{rva1}, one can
say that our proposed way to diagonalize the PPP-Hamiltonian in the selected
sub-Hilbert space is a very good approximation for appropriated parameters.
 
\section{polaronic states}

In this part, we consider the situation with one additional charge. We treat
explicitely the case of an additional electron but, the case of
the removal of one electron can be treated exactly in the same
way. We show that this problem can be describe, with some approximations, in terms of
quasi-particles which obey to simple effective Hamiltonian. For a
rigid lattice, we get a one dimensional tight binding
Hamiltonian. If one authorizes some distortions of the lattice around
the extra particle, we get at second order in the distortion
coordinates, a Holstein like model\cite{holstein}. In both cases, the
parameters of these one-electron models are related to the PPP one's.

In this work, we are not attempted to derive quantitative results. Our goal, based on semi-quantitative
results, is to open up a way
between a true many-body model given by the PPP-Hamiltonian and more simple
one-electron models as the Holstein's model for polaronic states. Because it is 
not possible to solve the PPP model and since the important physical
ingredients for an understanding of conjugated polymers are still not fully
recognized \cite{excitations}, the derivation of more effective models is
needed in order to get some physical insight. This work, and the very related one of
ref. \cite{pleutin}, goes in this direction.

For convenience, we choose in this part the simplest description for
the ground state given by the model I, using F, D and
Ct$^{-}_{1}$-LC. Since the model I already contains the most important local
constituents for the ground state wave function, namely the F and
Ct$^{-}_{1}$-LC, we believe the results
would not changed dramatically with a better description - by using the model
II or III. Then, if we define

\begin{equation}
\mid
n_{d},n_{ct}>=[C^{N-n_{ct}}_{n_{ct}}C^{N-2n_{ct}}_{n_{d}}]^{-1/2}\sum_{\{
y(i),z(j)\}}
Ct^{\dag}_{y(1)}...Ct^{\dag}_{y(n_{t})}D^{\dag}_{z(1)}\cdots
D^{\dag}_{z(n_{d})}\mid 0>
\end{equation} where the summation is over the
$C^{N-n_{ct}}_{n_{ct}}C^{N-2n_{ct}}_{n_{d}}$ possible configurations,
the ground state wave function is simply written as

\begin{equation}
\label{PF}
\mid
GS>=\sum^{N_{ct}}_{n_{ct}=0}a^{N_{ct}-n_{ct}}_{ct}{b^{n_{ct}}_{ct}}
\sqrt{C^{N_{ct}}_{n_{ct}}}\sum^{N-2n_{ct}}_{n_{d}=0}a^{N-{2n_{ct}}-n_{d}}_{c}
{b^{n_{d}}_{c}}\sqrt{C^{N-2n_{ct}}_{n_{d}}}\mid
n_{d},n_{ct}>
\end{equation}
where $N_{ct}={\bf E}(\frac{N-1}{3})$,
$a_{c}=\frac{(U-V)}{\sqrt{4\epsilon^{2}_{c}+(U-V)^{2}}}$,
$a^{2}_{c}+b^{2}_{c}=1$,
$a_{ct}=\frac{\sqrt{3}a^{2}_{d}t_{s}}{\sqrt{\epsilon^{2}_{t}+12a^{4}_{d}t^{2}
_{s}}}$
and $a^{2}_{ct}+b^{2}_{ct}=1$. $\epsilon$ is then the lowest
eigenvalue of the 2 by 2 matrix obtained from (\ref{3levels}) by
suppressing the effective level corresponding to the TT-LC\cite{pleutin}. For
a typical choice of parameters relevant for conjugated
polymers\cite{tavan}, the most probable LC is the F-LC
($a_{c}^{2}\simeq 0.98$ and $a_{ct}^{2}\simeq 0.25$); typical values
for the energies are given by $\epsilon_{c}\simeq -0.26eV$ and
$\epsilon \simeq -1.26eV$.

An additional charge disturbs the electronic cloud more or less strongly
depending on the system under consideration. It could be a local
distortion where the extra particle rearranges the system in short distances
to create around it what it is called polarization
cloud; this is the case for usual semi-conductors. On the contrary, it could be a
complete rearrangement of the system as for strongly correlated
systems\cite{fulde}. In our case, the first behaviour is concerned and a
quasi-particle picture is reached. 

\begin{figure}
\centerline{\psfig{figure=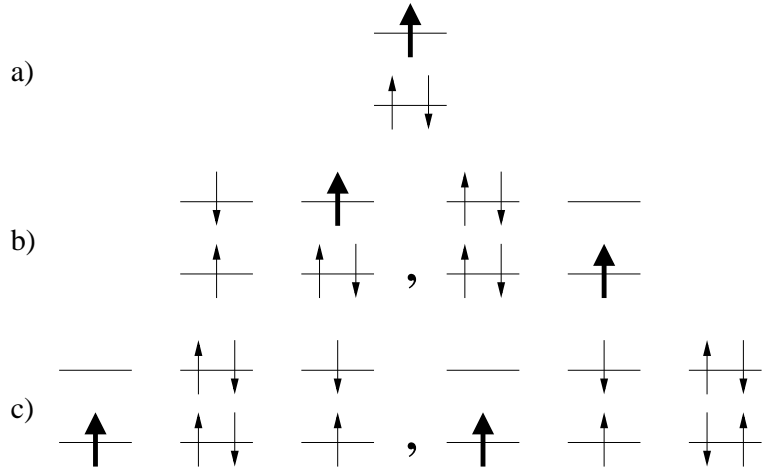,width=10cm}}
\caption{Examples of Charged-Local-Configurations (C-LC) extended over one
(a), two (b) and three (c) double bonds. The extra-particle is represented by
the thick arrow. The quasi-particle is identified with 
the P-LC (a), the lowest C-LC in energy.}
\label{LCP}
\end{figure}

We describe the perturbations caused by the
extra-particle - the polarization cloud - by introducing a new set of LC more or less extended,
which we call Charged Local Configurations (C-LC); the term "charged"
means that they contain explicitely the extra-particle. Some example
of C-LC, extended over one, two and three double bonds
are shown in figure (\ref{LCP}) where the extra-electron is
represented by the thick arrow. In the case of a "macroscopic" rearrangement of
the electronic structure - as it could be the case for strongly
correlated systems - the maximum extension of the relevant C-LC would
be of the order of the system size. In our case, this critical size is of the
order of some monomer units only.

All around these C-LC, we assume the electronic structure unchanged with
respect to the ground state;
therefore, we consider such charged configurations - strictly speaking, these
are linear combinations of electronic configurations but we adopt the
proposed terminology for convenience -
\begin{equation}
\label{excitations}
\begin{array}{c}
\mid \alpha_{n}>=\mid N_{L}>\otimes\mid C_{n}^{\alpha}>\otimes\mid N_{R}>\\
\mid \beta_{n,n+1}>=\mid N_{L}>\otimes\mid C_{n,n+1}^{\beta}>\otimes\mid N_{R}-1>\\
\mid \gamma_{n,n+1,n+2}>=\mid N_{L}>\otimes\mid C_{n,n+1,n+2}^{\gamma}>\otimes\mid N_{R}-2>
\end{array}
\end{equation}where, $\mid C_{n}^{\alpha}>$, $\mid
C_{n,n+1}^{\beta}>$ and $\mid C_{n,n+1,n+2}^{\gamma}>$ are some C-LC
extended over one, two and three nearest-neighbour double bonds respectively,
$\mid N_{L}>$ ($\mid N_{R}>$, $\mid N_{R}-1>$, $\mid N_{R}-2>$) is the
part on the left (right) of the C-LC, described in the same way as
$\mid GS>$. With this crude description, a C-LC acts as a dramatic
boundary which simply interrupts the chain: the system is separated into
two chains both described exactly as the ground state; the
boundary contains explicitely the extra-particle within a defined C-LC. The
more extended C-LC are inserted in the ground state in the same way as
(\ref{excitations}).

With our approximation, the energy of each charged configuration (as (\ref{excitations}))
is given by the addition of two different terms. The energy of the isolated
C-LC and the energy of the external parts to the left and to the right of
the C-LC. Since $\mid N_{L}>$ and $\mid N_{R}>$ are neutral, the external
parts don't interact via the Coulomb potential with the
extra-particle. However, the configurations (\ref{excitations}) must be improved for more quantitative
results. Indeed, in a better
description, because of the presence of the P-LC, the relative weight of the F, D and Ct$^{-}_{1}$-LC, controlled by
the coefficients $a_{c}$, $b_{c}$, $a_{ct}$ and $b_{ct}$ should depend on their
positions on the chain. Moreover, with an additional particle, the
electron-hole symmetry is broken. All the LC used in $\mid N_{L}>$ and $\mid N_{R}>$ are in the same sector
of symmetry - the proper one for the building of the ground state. This is the
case, for instance, of the Ct$^{-}_{1}$-LC where the charge transfer on the
right and on the left are of the same importance. In the presence of the P-LC
these two charge transfers are no longer equivalent; the symmetry is broken
and this implies a coulombic interaction between the P-LC and the external
parts. These effects, not considering in this work,
would certainly modify the polarisation cloud in a sensitive way. One can
say, in other
words, that the 'embedding" of the C-LC, due to the part $\mid
N_{L}>$ and $\mid N_{R}>$ of (\ref{excitations}), are not treated efficiently
in this work. We believe this is here the main point to be improved in
the future for more quantitative results.

For usual values of the PPP model, one kind of charged configurations is smaller in energy than the
other and in such way that a perturbative treatment is possible to do. These configurations are due to the C-LC, referred as the P-LC hereafter (P stands
for Particle), associated with the following creation operator

\begin{equation}
P^{\dag}_{n,\sigma}=A^{\dag}_{n,\sigma}F^{\dag}_{n}
\end{equation}and represented in figure (\ref{LCP}.a). The
extra-particle is immersed in the reference vacuum and gives the
following charged configurations

\begin{equation}
\label{particle}
\mid n>=\mid N_{L}>\otimes\mid P_{n}>\otimes\mid N_{R}>
\end{equation}where $n$ referred to the position of the P-LC, and with an energy given by
${\cal E}_{n} = \epsilon_{n}+(N^{r}+N^{l})\epsilon_{c} + (N^{r}+N^{l}
-3)\frac{\epsilon}{3}$, with $N^{r}+N^{l}=N-1$ and
$\epsilon_{n}=t_{d}+\frac{U}{2}+\frac{3V}{2}$, the energy of the
isolated P-LC. By comparing with (\ref{EGS}), we see that there is a loss of
intramonomer correlation energy and a loss of intermonomer fluctuation energy with respect to the ground state;
indeed, the additional electron occupies a site in which one cannot
place D and Ct$^{-}_{1}$-LC. This loss of
energy is more important for the more extended C-LC.

\begin{figure}
\centerline{\psfig{figure=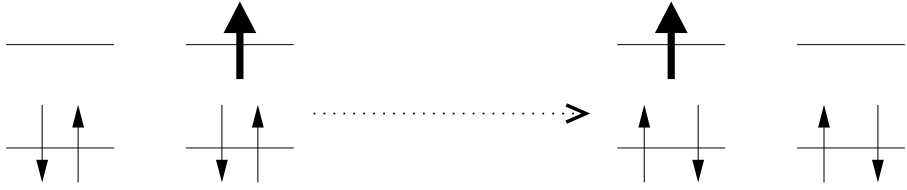,width=12cm}}
\caption{Nearest-neighbour hopping process for the P-LC assisted by the F-LC}
\label{nnhopping}
\end{figure}

\begin{figure}
\centerline{\psfig{figure=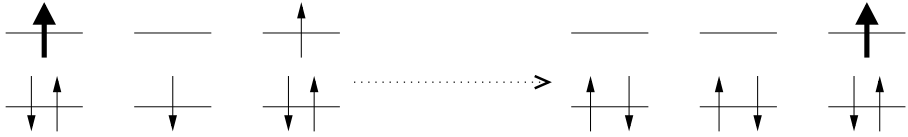,width=12cm}}
\caption{Next-nearest-neighbour hopping process for the P-LC assisted by the Ct$_{1}^{-}$-LC.}
\label{nnnhopping}
\end{figure}

In the following, we consider explicitely only the charged
configurations (\ref{particle}) since the effects of the other
charged configurations can be taken into account by perturbation. With our approximation, because of the n.n. hopping integral, the P-LC can hop 
on the lattice with the help of the F-LC or the Ct$^{-}_{1}$-LC. With the former, the P-LC can hop from site to site on the monomer lattice
(see figure \ref{nnhopping}).
\begin{equation}
\label{hopping}
<n \mid H_{ppp} \mid n \pm 1>=J=a_{c}^{2}b_{ct}^{2}\frac{t_{s}}{2}
\end{equation}In this expression, the product $a_{c}^{2}b_{ct}^{2}$ gives the
probability to find a F-LC in the wave function (\ref{PF}); the
factor $1/2$ in (\ref{hopping}) comes from our choice to work with the
monomer orbitals. Moreover, with our approximation, there exists also a
n.n.n. hopping process with
the help of the more extended GLC, the Ct$^{-}_{1}$-LC (see figure \ref{nnnhopping}).
\begin{equation}
\label{nexthopping}
<n \mid H_{ppp} \mid n \pm 2>=a_{ct}^{2}\frac{t_{s}}{4}
\end{equation}The additional factor 2 in the denominator comes from the
fact that only one term from the Ct$^{-}_{1}$-LC (see equation (\ref{ctLC}))
is involved during the transfer; the coefficient $a_{ct}^{2}$ gives the
probability to find a Ct$^{-}_{1}$-LC in the ground state wave function
(\ref{PF}). The n.n.n. transfer is of course less
important than the n.n. one's. With the values for the parameters we use
here, the values of these two hopping processes differ by one order of
magnitude. Therefore, we neglect the n.n.n. effective hopping term in this work.

The extra-particle (P-LC) can be dressed by perturbation. Some effects of the other
C-LC appear then in renormalized energy and n.n. hopping term for the
extra-particle. This dressing of the P-LC can be simply done by a second order
perturbative treatment, giving, in one hand, the so-called polarization energy
\begin{equation}
\label{epolarisation}
\epsilon_{p}=\sum_{\delta}\frac{t_{\delta}^{2}}{{\cal E}_{n}-{\cal E}_{\delta}}
\end{equation}and, in the other hand, some corrections for the n.n. hopping integral $J$

\begin{equation}
\label{effhopping}
J_{eff}=\sum_{\delta, \delta^{'}}t_{\delta}t_{\delta^{'}}(\frac{1}{{\cal E}_{n}-{\cal E}_{\delta}}+\frac{1}{{\cal E}_{n}-{\cal E}_{\delta^{'}}})
\end{equation}In these expressions $t_{\delta}$ and $t_{\delta^{'}}$ are
some interacting terms between the P-LC and other C-LC. The inequalities $\mid
\frac{t_{\delta/\delta^{'}}}{{\cal E}_{n}-{\cal E}_{\delta/\delta^{'}}} \mid
<<1$ are respected for the values of the parameters we use which guaranty the
relevance of a perturbative treatment. After that, we
have reached a quasi-particle picture, the quasi-particle being represented by the P-LC.

\begin{figure}
\centerline{\psfig{figure=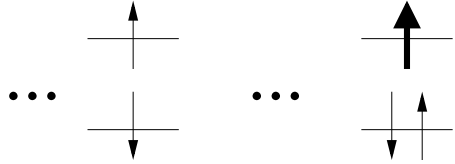,width=6cm}}
\caption{Example of Local Configurations including long range
polarization effects}
\label{LClongrangepolar}
\end{figure}

In principle, many C-LC give some contributions to the perturbative series
(\ref{epolarisation}) and (\ref{effhopping}). However, because the states
(\ref{excitations}) ignore many effects due to an inappropriate embedding, as
we already mentioned, we believe it is not useful to carry out the 
full calculation. Consequently, we do here a simplified
treatment for the dressing of the extra-particle which we believe contains
anyhow the most important contributions to (\ref{epolarisation}) and (\ref{effhopping}). This simplified 
treatment consists to consider the C-LC not embedded in the ground state defined by
(\ref{PF}) but in a simplified vacuum made of only F-LC. Since, the
F-LC is the very most important LC in the ground state (\ref{PF}), we believe
this simplified treatment sufficient to capture the most important parts of the
polarisation energy and the effective hopping term. Moreover, among the
remaining charged configurations only a few are incorporated in the
perturbative treatment; they are shown in the figure (\ref{LCP}.b). By this
last simplification we neglect all the C-LC shown in the figure
(\ref{LClongrangepolar}) which take into account some long range polarisation
effects; these C-LC are numerous but their total effect on
(\ref{epolarisation}) and (\ref{effhopping}) are small
and they don't participate sensitively to the binding energy of the polaron
state which is the main quantity we are looking for here. With this treatment, the corrections for the hopping term remain always negligible
in the range of parameters of interest; we will therefore neglect
these last corrections, $J_{eff}$.

After the dressing operation, we obtain formally a
one particle like problem with two characteristic energy terms
$E_{n}$, the site energy of the additional 'electron' with respect to
the ground state, and $J$,
the hopping term, which are functions of the PPP parameters:
$E_{n}=\epsilon_{n}-\epsilon_{c}-\frac{2}{3}\epsilon+\epsilon_{p}$ and
$J=\frac{t_{s}}{2}$.
If we suppose a rigid lattice, the problem can obviously be
diagonalized, giving a band centred at $E_{n}$, with a bandwidth
of $4\mid J \mid$. In the case of the SSH Hamiltonian\cite{ssh} and
by neglecting $\epsilon$ and $\epsilon_{p}$, the bottom of the
band is given by $E_{n}=\mid t_{d}-t_{s}\mid$, the exact result;
with inclusion of these corrective terms, this energy
becomes slightly overestimated. The effective mass associated with the P-LC is given by
$m^{*}\simeq\frac{\hbar^{2}}{a^{2}}\frac{1}{t_{s}}$ ($a$ is the unit cell
length) which is of course higher than the effective mass of a free
particle on the bottom of the conduction band. With the Coulomb term,
$E_{n}$ increases and $m^{*}$ stays unchanged. In conclusion, for a rigid
lattice, we have reached a simple tight 
binding Hamiltonian - the so-called H\"uckel model. Last, one may say that
such approach is quite close in spirit of a recent work of J. Grafenstein et
al., where an effective tight-binding model is derived at ab-initio level by
means of an incremental method \cite{grafenstein}.

\begin{figure}
\centerline{\psfig{figure=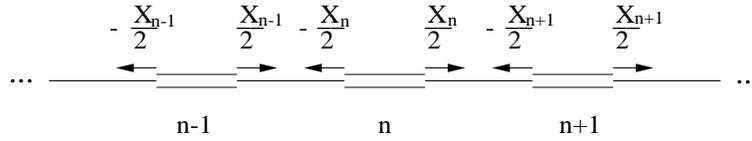,width=10cm}}
\caption{Model for the lattice deformation adopted in this work.}
\label{deformation}
\end{figure}

Now we allow a relaxation of the
lattice. For simplicity, we choose a
model displacement where the two 'atoms' of the same double
bond move
with the same amplitude $\frac{\mid x_{n} \mid}{2}$ but in opposite directions (cf. figure (\ref{deformation})).
The two parameters
$E_{n}$ and $J$ depend now on the lattice
coordinates via mainly the linear dependence of the two hopping terms,
$t_{d}(x_{n})=t_{d}-\alpha x_{n}$ and $t_{s}(x_{n},x_{n+1})=t_{s}+\alpha
(\frac{x_{n}}{2}+\frac{x_{n+1}}{2})$. On the
contrary, the
coulombic terms remain almost unchanged after a small displacement.
The contributions due to these displacements to $E_n$ and $J$ are small so we
make a linear expansion with respect to $\{x_{n}\}$ of these two quantities

\begin{equation}
E({x_{n}})=E_{n}-\alpha (a_{0}x_{n} + a_{1}(x_{n+1}+x_{n-1}))
\end{equation}

\begin{equation}
J({x_{n}})=J-\alpha b_{0}(x_{n}+x_{n+1})
\end{equation}
where $a_{0}$, $a_{1}$ and $b_{0}$ are functions of PPP parameters
and $\alpha$
is the electron-phonon interaction term\cite{ssh}. The extra
elastic constraint of the dimerized chain due to the lattice
relaxation in presence of an additional charge is expressed as

\begin{equation}
E_{el}=\frac{1}{2}\sum_{n}K_{eq}[x^{2}_{n}+(\frac{x_{n}}{2}
+\frac{x_{n-1}}{2})^2]
\end{equation}where $K_{eq}$, the spring constant, is defined
relatively to the dimerized equilibrium structure.

The coefficients $C_{n}$ of the Holstein polaron wave function\cite{holstein},
$\mid \Psi_{p}>=\sum_{n}C_{n}({x_{n}})\mid n>
$, are determined by minimization of the corresponding total energy,
$E_{T}(\{x_{n}\})$, with respect to the lattice coordinate $x_{n}$.
At the second order in $x_{n}$ and taking into account that
$\frac{\alpha}{K_{eq}}\sim0.1\AA$ in conjugated polymers \cite{revue}, we obtain the
characteristic equations of the molecular Holstein's model

\begin{equation}
\label{holstein}
[Fx_{n}-2J-\epsilon]C_{n}+JC_{n+1}+JC_{n-1}=0
\end{equation}

\begin{equation}
\label{relaxation}
x_{n}k=F\mid C_{n}\mid ^{2}
\end{equation}
where the coefficients are expressed in function of the PPP
parameters: $F=(a_{0}+2a_{1}+4b_{0})\alpha$, $k=2K_{eq}$ and
$J=\frac{t_{s}}{2}$. By injecting (\ref{relaxation}) into
(\ref{holstein}) we obtain the non-linear Schr\"odinger
equation which gives the coefficients of the wave function; the second
equation connects in a simple manner these coefficients and the lattice
deformation. The analytical solution of these two equations in the
continuum limit\cite{holstein}, valid for the "large" polaron case, gives
the well known polaronic wave function
$C_{n}=\frac{\gamma}{\eta}\mbox{sech}(\gamma(n-n_{0}))$
with $E_{b}=\frac{F^{2}}{2k}$, $\eta^{2}=\frac{E_{b}}{J}$ and
$\gamma^{2}=\frac{\eta^{2}}{2}$; the polaron state is localized around
$n_{0}$, an undetermined quantity because of the translational
invariance of the system. The associated binding energy of the
polaron state is given by $E_{p}=\frac{E^{2}_{b}}{12J}$.

We evaluate these quantities for several choices of parameters by the
following sequence of calculations. First we
optimize the dimerized geometry referring to a spring constant, K,
relative to a hypothetical undimerized geometry\cite{ssh}; then, we
evaluate $K_{eq}$, calculating the
second derivative of $E_{T}$ with respect to the dimerization
coordinate at the geometrical equilibrium. Second, we solve the
equations (\ref{holstein}) and (\ref{relaxation}).

In the continuum version of the SSH Hamiltonian limit\cite{campbell},
analytical expressions have been given. Our results always overestimate the
reported values. For example, with $t_{0}=2.5 eV$, $\alpha=4.1
eV\AA^{-1}$ and $K=21 eV\AA^{-2}$, we get $E_{p}=0.11eV$ in place of
$0.064eV$. In the same manner, our method also overestimate the value of
the dimerization. These overestimations occur naturally from our
starting point which relies on a molecular description. Besides it has
been shown that the SSH Hamiltonian is never equivalent to the Holstein's
model for the dimerized linear chain\cite{campbell}, so that the
approximations of our model cannot be expected to lead to a good
agreement in this case.
However our approximations will cope better when the Coulomb interaction
is taken into account; then the energies of charge fluctuation
components decrease with their extensions, due to the long range
part of the potential. This fact is in favour of our
approximation. Furthermore the value of $K$ used in this example
is the appropriated one for the SSH Hamiltonian\cite{ssh},
but seems not to be in agreement with the experimental results
obtained for small oligomers\cite{revue}. An higher
value must be taken, favouring again our description.

If one adds the Ohno potential, the binding energy
decreases: as example, for the same choice of
parameters and $U=11.16eV$, we get $E_{p}=0.091eV$. Finally, taking
the same parameters but with a more appropriate value for $K$, $K=41
eV\AA^{-2}$, we get a reasonable equilibrium
geometry characterized by $r_{d}=1.33\AA$ and $r_{s}=1.47\AA$. Moreover we
get the following values
$F\simeq9.5eV\AA^{-1}$, $J\simeq1.1eV$, $k\simeq78eV\AA^{-2}$ and the
binding energy for the polaron decreases to $E_{p}\simeq0.025eV$. In any case, 
we stay around traditionally adopted values.

Before closing this section, note that with such a low binding energy,
expected for conjugated polymers, the quantum fluctuations of the lattice
should be explicitely considered. However, it is for the moment totally
hopeless to introduce additional bosonic variables in the full PPP Hamiltonian.

\section{conclusion}

In conclusion, we have proposed a simplified treatment of the PPP
Hamiltonian which is typically a diagonalization of this Hamiltonian
in a restricted Hilbert space. The method adopted, using monomer
orbitals, is a natural way to bridge the gap between small cluster and
polymer calculations\cite{chandross,pleutin}. The ground state is composed of
intermonomer nearest-neighbour fluctuation components
in the background of coupled electrons by pairs localized on
monomers. Comparisons with DMRG results for the extended Peierls-Hubbard model 
show satisfactory agreements considering the simplicity of our proposed wave
functions. The electronic excitations are then described as local perturbations
moving in this "vacuum". For an appropriate set of parameters,
this description gives rather good values for the dimerisation
and for the energies of the excited states active in one photon
spectroscopy\cite{pleutin}. In the doped case (2N+1 particles) studied here,
following the adiabatic scheme proposed by Holstein\cite{holstein}, we show
that our model leads naturally to a Holstein's polaron like problem. However
our description differs drastically from the Holstein's polaron image in the
sense that it is able to describe the behaviour of a strong correlated (2N+1)
particle state whereas the Holstein's model considers
only the additional particle in interaction with a deformable
medium. The obtained binding energy of the polaron is of the correct order of
magnitude.

Some improvements are suitable concerning, first, the ground state where more
extended GLC must be considered in order to reproduce more accurately the
delocalization proper to $\pi$-systems. In other hand, variational calculations
based on the very same ideas are possible \cite{rva1,rva2,mp}. For the doped case, we believe the first
point would be to improve the description of the vacuum in presence of the
extra-particle. Even if it is difficult to test our derivation in part owing
to a lack of accurate calculations including correlation effects, we
think that our formulation keeps the essential behaviour of
the considered physical phenomenon and believe that it could be useful for
future more advanced studies in part because of its relative conceptual
simplicity and of its ability to give analytical expressions.
For example, the behaviour of polaron states in
presence of a strong electric field \cite{bussac}, which corresponds to a common
situation in electroluminescence studies, would be to consider taking
into account the effects of the strongly correlated N particle
system.

\begin{acknowledgements}
We wish to thank E. Jeckelmann for giving us DMRG results prior to
publication. S.P acknowledges support from the European Commission through the
TMR network contract ERBFNRX-CT96-0079 (QUCEX).
\end{acknowledgements}

\begin{table}
\label{ssh}
\caption{Percentage of the exact energy obtained with the different models
studied here for the S.S.H. Hamiltonian. The model I contains the F, D and
Ct$_{1}^{-}$-LC; the TT-LC are added for the model II and the whole set of LC
shown in figure (\ref{LCGS}) are considered for the model III.}
\begin{tabular}{cccc}
&model I&model II&model III\\
$x=0.$&91.6$\%$&92.1$\%$&92.7$\%$\\
$x=0.15$&96.7$\%$&96.9$\%$&97.2$\%$
\end{tabular}
\end{table}

\begin{table}
\label{uv}
\caption{Energy per unit cell for an infinite lattice obtained with the three successive
approximations (model I, II, III) and DMRG calculations for the extended
Peierls-Hubbard model with $U=4t$ and $V=t$; in the case of DMRG, the energies per
unit cell are obtained from extrapolation of large cluster calculations up to
400 sites.}
\begin{tabular}{ccccc}
&model I&model II&model III&DMRG\\
$x=0.05$&0.373311&0.369217&0.366820&0.313599\\
$x=0.15$&0.306566&0.304317&0.303046&0.270381\\
$x=0.25$&0.236707&0.235678&0.235022&0.213969\\
$x=0.75$&-0.160370&-0.159835&-0.159840&-0.164925
\end{tabular}
\end{table}
\end{document}